\documentclass[10pt]{article}

\usepackage{amsfonts}
\usepackage{amsmath, graphicx, epsfig,psfrag}

\usepackage[usenames]{color}
\usepackage{epstopdf}

\usepackage{undertilde}

\newcommand{\ez}{\mathbb{E}}
\newcommand{\pz}{\mathbb{P}}

\numberwithin{equation}{section}

\begin{document}

\baselineskip20truept \pagestyle{plain}

\pagenumbering{arabic}

\title{The basic reproduction number, $R_0$, in structured populations}
\author{Peter Neal (Lancaster University)\footnote{Corresponding author; Department of Mathematics and Statistics, Fylde College, Lancaster University, Lancaster, LA1 4YF, United Kingdom. email: p.neal@lancaster.ac.uk} and Thitiya Theparod (Mahasarakham University)\footnote{Statistics and Applied Statistics Research Unit, the Department of Mathematics, Mahasarakham University, Khamriang Sub-District, Kantarawichai District, Maha Sarakham 44150, Thailand}}
\maketitle



\begin{abstract}
In this paper, we provide a straightforward approach to defining and deriving the key epidemiological quantity, the basic reproduction number, $R_0$, for Markovian epidemics in structured populations. The methodology derived is applicable to, and demonstrated on, both $SIR$ and $SIS$ epidemics and allows for population as well as epidemic dynamics. The approach taken is to consider the epidemic process as a multitype process by identifying and classifying the different types of infectious units along with the infections from, and the transitions between, infectious units. For the household model, we show that our expression for $R_0$ agrees with earlier work despite the alternative nature of the construction of the mean reproductive matrix, and hence, the basic reproduction number.
\end{abstract}

{\sc Keywords:} Basic reproduction number; $SIR$ epidemics; $SIS$ epidemics; household epidemic model; sexually transmitted diseases. 

\section{Introduction} \label{S:intro}

The basic reproduction number, $R_0$, is a key summary in infectious disease modelling being defined as the expected number of individuals infected by a typical individual in a completely susceptible population. This definition of $R_0$ is straightforward to define and compute in a homogeneous population consisting of a single type of infective (homogeneous behaviour) and with uniform random mixing of infectives (homogeneous mixing). This yields the celebrated threshold theorem, see for example, \cite{Whittle55}, for the epidemic with a non-zero probability of a major epidemic outbreak if and only if $R_0 >1$.

The extension of the definition $R_0$ to non-homogeneous populations is non-trivial. Important work in this direction includes \cite{Diekmann_etal} which considers heterogeneous populations consisting of multiple types of infectives and \cite{Pellis_etal}, \cite{Ball_etal} which consider heterogeneity in mixing through population structure. Specifically \cite{Diekmann_etal} defines for a population consisting of $K$ types of infectives, the $K \times K$ mean reproduction matrix $\mathbf{M}$ (also known as the next-generation-matrix), where $M_{ij}$ denotes the mean number of infectives of type $j$ generated by a typical type $i$ infective during its infectious period. Then $R_0$ is defined as the Perron-Frobenius (dominant) eigenvalue of $\mathbf{M}$. By contrast \cite{Pellis_etal} and \cite{Ball_etal} focus on a household epidemic model with a single type of infective and consider a branching process approximation for the initial stages of the epidemic process, see, for example, \cite{Whittle55}, \cite{Ball_Donnelly} and \cite{BMST}. \cite{Pellis_etal} consider the asymptotic growth rate of the epidemic on a generational basis using an embedded Galton-Watson branching process and define $R_0$ to be
\begin{eqnarray} \label{eq:intro:1} R_0 = \lim_{n \rightarrow \infty} \ez [X_n]^{1/n}, \end{eqnarray}
where $X_n$ is the number of infectives in the $n^{th}$ generation of the epidemic. Given that the mean reproduction matrix $\mathbf{M}$ represents the mean number of infectives generated by an infective in the next generation, we observe that the computation of $R_0$ in both cases is defined in terms of the generational growth rate of the epidemic.

The current work applies the approach of \cite{Diekmann_etal} to Markovian epidemics in structured populations and thus assumes individuals have exponentially distributed infectious periods. The key idea is that in structured populations we can define infectives by the type of infectious unit to which they belong, where for many models the number of type of infectious units is relatively small and easy to classify. By characterising an infective by the type of infectious unit they originate in (belong to at the point of infection) and considering the possible events involving the infectious unit, we can write down a simple recursive equation for the mean reproduction matrix $\mathbf{M}$. Then as in \cite{Diekmann_etal} we can simply define $R_0$ to be the Perron-Frobenius eigenvalue of $\mathbf{M}$.

Our approach is similar to \cite{LKD15}, who also consider classifying infectives by the type of infectious unit to which they belong in an dynamic $SI$ sexually transmitted disease model which is similar to the $SIS$ model studied in Section \ref{S:example:sex}. The modelling in \cite{LKD15} is presented in a deterministic framework with \cite{Lashari_Trapman} considering the model from a stochastic perspective. The key difference to \cite{LKD15} is that we work with the embedded discrete Markov process of the transition events rather than the continuous time Markov rate matrices. The advantages of studying the discretised process is that it is easier to incorporate both local (within-infectious unit) infections and global (creation of new infectious units) infections and in generalisations of the construction of $\mathbf{M}$ beyond exponential infectious periods, see Section \ref{S:example:gcm} and Appendix \ref{App:rank}. Moreover, we present the approach in a general framework which easily incorporates both $SIR$ and $SIS$ epidemic models and allows for population as well as epidemic dynamics.

The remainder of the paper is structured as follows. In Section \ref{S:setup} we define the generic epidemic model which we consider along with the derivation of $\mathbf{M}$ and $R_0$. To assist with the understanding we illustrate with an $SIR$ household epidemic model (\cite{BMST}, \cite{Pellis_etal} and \cite{Ball_etal}). In Section \ref{S:Example}, we detail the computing of  $\mathbf{M}$ and $R_0$, for the $SIR$ household epidemic model (Section \ref{S:example:house}), an $SIS$ sexually transmitted disease model (Section \ref{S:example:sex}), see, for example, \cite{Kret}, \cite{LKD15} and \cite{Lashari_Trapman} and the great circle $SIR$ epidemic model (Section \ref{S:example:gcm}), see \cite{BMST}, \cite{Ball_Neal02} and \cite{Ball_Neal03}. In Section \ref{S:example:house} we show that the computed $R_0$ agrees with $R_0^g$ obtained in \cite{Ball_etal}.

%
%
%

\section{Model setup} \label{S:setup}

In this Section we characterise the key elements of the modelling. In order to keep the results as general as possible, we present a generic description of the model before illustrating with examples to make the more abstract concepts more concrete.

We assume that the population consists of $M$ types of individuals and for 
illustrative purposes we will assume that $M=1$. We allow individuals to be grouped together in local units and a unit is said to be an infectious unit if it contains at least one infectious individual. The local units might be static (remain fixed through time) or dynamic (varying over time). We assume that there are $K$ states that local infectious units can be in. Note that different local units may permit different local infectious unit states. Finally, we assume that all dynamics within the population and epidemic are Markovian. That is, for any infectious unit there is an exponential waiting time until the next event involving the infectious unit and no changes occur in the infectious unit between events. We assume that there are three types of events which take place within the population with regards the epidemic process. These are:-
\begin{enumerate}
\item {\bf Global infections}. These are infections where the individual contacted is chosen uniformly at random from a specified type of individual in the population. If the population consists of one single type of individual then the individual is chosen uniformly at random from the whole population. It is assumed that the number of individuals of each type are large, so that in the early stages of the epidemic with probability tending to 1, a global infectious contact is with a susceptible individual, and thus, results in an infection.
\item {\bf Local transitions}.  These are transitions which affect an infectious unit. These transitions can include infection within an infectious unit leading to a change of state of an infectious unit or an infectious individual moving to a different type.
\item {\bf Recovery}. An infectious individual recovers from the disease and is no longer able to infect individuals within their current infectious episode. Given that we allow for $SIR$ and $SIS$ epidemic dynamics, a given individual may have at most one, or possibly many, infectious episodes depending upon the disease dynamics.
\end{enumerate}

\subsection{$SIR$ Household example}

An example of an epidemic model which satisfies the above setup is the $SIR$ household epidemic model with exponential infectious periods. We illustrate assuming that all households are of size $h >1$ with the extension to allow for different size households trivial. An individual, whilst infectious, makes global contacts at the points of a homogeneous Poisson point process with rate $\lambda_G$ with the individual contacted chosen uniformly at random from the entire population and local contacts at the points of a homogeneous Poisson point process with rate $(h-1) \lambda_L$ with the individual contacted chosen uniformly at random from the remaining $h-1$ individuals in the infectives household. It is assumed that the local and global contacts are independent. Note that an infective makes contact with a given individual in their household at rate $\lambda_L$. Infectives have independent and identically distributed exponential infectious periods with mean $1/\gamma$ corresponding to infectives recovering at rate $\gamma$. In this case $M=1$ although we could extend to a multitype household model, see \cite{Ball_Lyne}. Infectious units correspond to households containing at least one infective and we classify households by the number of susceptibles and infectives they contain. Therefore the possible infectious states of a household are $\{(a,b); b=1,2,\ldots,h; a=0,1,\ldots, h-b \}$, where $a$ and $b$ denote the number of susceptibles and the number of infectives in the household, respectively. Thus there are $K = h (h+1)/2$ states. A global infection with a previously uninfected household results in the creation of a new infectious unit in state $(h-1,1)$. A local infection in a household in state $(a,b)$ results in the household moving to state $(a-1,b+1)$, whilst a recovery in a household in state $(a,b)$ results in the household moving to state $(a,b-1)$, and no longer being an infectious unit if $b=1$.

\subsection{Definition of $R_0$}

We define
 $R_0$ as the maximal eigenvalue of the mean reproduction matrix $\mathbf{M}$, where $\mathbf{M}$ is a $K \times K$ matrix with $m_{ij}$ denoting the mean number of state $j$ infectious units generated by an infective who enters the infectious state as a member of a state $i$ infectious unit. This definition of $R_0$ is consistent with earlier work on computing the basic reproduction number in heterogeneous populations with multiple types of infectives, see for example, \cite{Diekmann_etal}. A key point to note is that $\mathbf{M}$ will capture only the infections made by a specific infective rather than all the infections made by the infectious unit to which they belong.

Linking the mean reproduction matrix $\mathbf{M}$ back to the $SIR$ household example. An individual will be classed as a state $(a,b)$ individual if the event which leads to the individual becoming infected results in the infectious unit (household) entering state $(a,b)$. An infective will generate a state $(c,d)$ individual, if they are infectious in an infectious unit in state $(c+1,d-1)$ and the infective is responsible for infecting one of the susceptibles in the household. Note that if $d \geq 3$, the infectious unit can transit from state $(c+1,d-1)$ to state $(c,d)$ without the infective in question having made the infection. 

\subsection{Construction of $\mathbf{M}$}


Consider an infective belonging to an infectious unit of state $i$ ($i=1,2,\ldots,K$). Suppose that there are $n_i$ events which can occur to an infectious unit in state $i$. Let $q_{il}$ $(i=1,2,\ldots,K;l=1,2,\ldots,n_i)$ denote the probability that a type $l$ event occurs in the infectious unit. Let $a_{il}$ $(i=1,2,\ldots,K;l=1,2,\ldots,n_i)$ denote the state of the infectious unit following the type $l$ event with $a_{il}= 0$ if the infective recovers and so no longer infectious. Let $Y_{ilj}$ $(i,j=1,2,\ldots,K;l=1,2,\ldots,n_i)$ denote the total number of type $j$ infectious units generated by an infective who undergoes a type $l$ event
 with $\mu_{ilj} = E[Y_{ilj}]$ and $\mathbf{Y}_{il} = (Y_{il1}, Y_{il2}, \ldots, Y_{ilK})$.



For $i,j=1,2,\ldots,K$, let
\begin{eqnarray} \label{eq:M:prep:1} p_{ij} = \sum_{l=1}^{n_i} 1_{\{a_{il} =j \}} q_{il}, \end{eqnarray} the probability that an infective belonging to a state $i$ infectious unit moves to a state $j$ infectious unit. Note that typically $\sum_{j=1}^K p_{ij} <1$ as there is the possibility of the infective recovering from the disease and let $p_{i0} = 1- \sum_{j=1}^K p_{ij}$, the probability a type $i$ infective recovers from the disease. For $i,j=1,2,\ldots,K$, let
\begin{eqnarray} \label{eq:M:prep:2} \phi_{ij} = \sum_{l=1}^{n_i} q_{il} \mu_{ilj}, \end{eqnarray} the mean number of state $j$ infectious units generated by an event directly involving an infective in a state $i$ infectious unit. It follows using the theorem of total probability and the linearity of expectation that
\begin{eqnarray} \label{eq:M:prep:3}
m_{ij} &=& \sum_{l=1}^{n_i} q_{il} \ez[\mbox{State $j$ infectious units generated} | \mbox{Type $l$ event}] \nonumber \\
&=& \sum_{l=1}^{n_i} q_{il} \left\{ \mu_{ilj} + m_{a_{il} j} \right\} \nonumber \\
&=& \sum_{l=1}^{n_i} q_{il} \mu_{ilj} + \sum_{l=1}^{n_i} q_{il}  m_{a_{il} j} \nonumber \\
&=& \phi_{ij} + \sum_{i=1}^{n_i} q_{il} \left\{ \sum_{k=1}^{K} 1_{\{a_{il} = k \}} m_{kj} \right\} \nonumber \\
&=& \phi_{ij} + \sum_{k=1}^{K}   \left\{  \sum_{i=1}^{n_i} q_{il} 1_{\{a_{il} = k \}} \right\} m_{kj}  \nonumber \\
&=& \phi_{ij} +  \sum_{k=1}^{K} p_{ik} m_{kj}, \end{eqnarray}
where for $j=1,2,\ldots,K$, let $m_{0j} =0$, that is, a recovered individual makes no infections. 

Letting $\Phi = (\phi_{ij})$ and $\mathbf{P} = (p_{ij})$ be $K \times K$ matrices, we can express \eqref{eq:M:prep:3} in matrix notation as
\begin{eqnarray} \label{eq:M:1} \mathbf{M} = \Phi + \mathbf{P} \mathbf{M}. \end{eqnarray}
Rearranging \eqref{eq:M:1}, with $\mathbf{I}$ denote the $K \times K$ identity matrix, we have that
\begin{eqnarray} \label{eq:M:2} \mathbf{M} = \left(\sum_{n=0}^\infty \mathbf{P}^n \right) \Phi = (\mathbf{I} - \mathbf{P})^{-1} \Phi. \end{eqnarray}
Since individuals recover from the disease,  $\mathbf{P}$ is a substochastic matrix with at least some rows summing to less than 1. Thus the Perron-Frobenius theorem gives that $\mathbf{P}^n \rightarrow \mathbf{0}$ as $n \rightarrow \infty$.

The definition of an event, and hence, $\mathbf{Y}_{il}$ can be accomplished in a variety of ways. In this paper, we typically take an event to coincide with a change in the type of infectious unit to which an infective belongs and we take account of the (mean) number of global infections an infective makes in a type $i$ infectious unit before transitioning to a new type of infectious unit. In this way $p_{ii} =0$ $(i=1,2,\ldots,K)$. An alternative approach is to define an event to be any global infection, local infection or recovery within their infectious unit. In this case nothing occurs between events and $\mathbf{Y}_{il}$ is the number of infectious units generated by the type $l$ event. In Section \ref{S:example:sex}, we present both constructions for an $SIS$ sexually transmitted disease model.  

\subsection{Comments on $\mathbf{M}$}

We make a couple of observations concerning the construction of $\mathbf{M}$.

In the $SIR$ household epidemic model, we can reduce the number of states to $h(h+1)/2 - (h-1)$ by noting that in households with 0 susceptibles, no local infections can occur and thus infectives can only make global infections acting independently. Therefore we can subsume the states $(0,1), (0,2), \ldots, (0,h)$ into a single state, $(0,1)^\ast$ say, with a local infection in households with 1 susceptible resulting in the household moving to state $(0,1)^\ast$, see for, example \cite{Neal16}.

For $SIR$ epidemics, there will typically be a natural ordering of infectious unit types such that we can order the types with only transitions of infectious unit from type $i$ to type $j$ ($i < j$) being possible. For example, with household epidemics we can order the types such that type $(a,b)$ is said to be less than type $(c,d)$, if $a >c$, or if $a=c$, and $b>d$. In such cases $\mathbf{P}$ is an upper triangular matrix and if the main diagonal of $\mathbf{P}$ is $\mathbf{0}$ then there exists $n_0 \leq K$, such that for all $n > n_0$, $\mathbf{P}^n = \mathbf{0}$. Then
\begin{eqnarray} \label{eq:M:3} \mathbf{M} = (\mathbf{I} - \mathbf{P})^{-1} \Phi =\left(\sum_{n=0}^{n_0} \mathbf{P}^n \right) \Phi, \end{eqnarray}
and we can compute $\mathbf{M}$ without requiring matrix inversion.

\section{Examples} \label{S:Example}

In this Section we show how $\mathbf{M}$ is constructed for three different models; the household SIR epidemic model (Section \ref{S:example:house}), an SIS sexually transmitted disease (Section \ref{S:example:sex}) and the great circle SIR epidemic model  (Section \ref{S:example:gcm}).

\subsection{$SIR$ Household epidemic model} \label{S:example:house}

We illustrate the computation of $R_0$ in a population with households of size $h$. As noted in Section \ref{S:setup}, we can summarise the epidemic process using $K = h (h+1)/2 - (h-1)$ states by amalgamating states $(0,1), (0,2), \ldots, (0,h)$ into the state $(0,1)^\ast$. We use the labellings $\{(0,1)^\ast, (a,b); a,b=1,2,\ldots,h, (a+b) \leq h\}$ rather than $1,2, \ldots, K$ to define the mean reproduction matrix.


We construct $\mathbf{M}$ by first considering the local transitions (infections and recoveries) which occur within a household. Therefore for an individual in state $(a,b)$, the non-zero transitions are
\begin{eqnarray}
p_{(a,b),(a-1,b+1)} &=& \frac{a b \lambda_L }{b (a \lambda_L+ \gamma)}  \hspace{0.5cm} \mbox{if $a>1$} \nonumber \\
p_{(a,b),(0,1)^\ast} &=&  \frac{a b\lambda_L}{b (a \lambda_L + \gamma)} \hspace{0.5cm} \mbox{if $a=1$} \label{eq:house:E:0}  \\
p_{(a,b),(a,b-1)}&=&  \frac{(b-1) \gamma}{b (a \lambda_L + \gamma)}. \nonumber
 \end{eqnarray}
Note that the probability that the next event that the individual of interest recovers is $\gamma/\{ b (a \lambda_L + \gamma)\}$ and an individual only leaves state $(0,1)^\ast$ through recovery. Therefore the transition probabilities in \eqref{eq:house:E:0} define the substochastic matrix $\mathbf{P}$. The time that a household spends in state $(a,b)$ is exponentially distributed with rate $b (a \lambda_L + \gamma)$. Therefore, since infectives are making infectious contacts at the points of a homogeneous Poisson point process with rate $\lambda_G$, the mean number of global contacts made by an infective, whilst the household is in state $(a,b)$, is $\lambda_G/\{ b (a \lambda_L + \gamma)\}$  with all global contacts resulting in an $(h-1,1)$ infectious unit. This gives the non-zero entries of $\Phi$ to be
\begin{eqnarray*}
\phi_{(a,b),(a-1,b+1)} &=& \frac{a \lambda_L }{b (a \lambda_L + \gamma)} = \frac{p_{(a,b),(a-1,b+1)}}{b}  \hspace{0.5cm} \mbox{if $a>1$} \\
\phi_{(a,b),(0,1)^\ast} &=&  \frac{\lambda_L}{b (a \lambda_L + \gamma)} = \frac{p_{(a,b),(0,1)^\ast}}{b} \hspace{1.05cm} \mbox{if $a=1$} \\
\phi_{(a,b),(h-1,1)} &=&  \frac{\lambda_G}{b (a \lambda_L + \gamma)}   \end{eqnarray*}
Note that the probability that the infective of interest is responsible for a given local infection in a household in state $(a,b)$ is simply $1/b$.

In a population of households of size $3$ with the states ordered $(2,1)$, $(1,2)$, $(1,1)$ and $(0,1)^\ast$, we have that
\begin{eqnarray} \label{eq:house:E:1}
\mathbf{P} &=& \begin{pmatrix} 0 &  \frac{2 \lambda_L}{2 \lambda_L + \gamma} & 0 & 0 \\
 0&  0 &  \frac{\gamma}{2 (\lambda_L  + \gamma)} &    \frac{2\lambda_L}{2 (\lambda_L + \gamma)} \\
 0 & 0 &0& \frac{\lambda_L}{\lambda_L  + \gamma} \\
 0 & 0 & 0 & 0 \end{pmatrix} \end{eqnarray}
and
\begin{eqnarray} \label{eq:house:E:2}
\Phi &=& \begin{pmatrix} \frac{\lambda_G}{2 \lambda_L  + \gamma} &  \frac{2 \lambda_L}{2 \lambda_L + \gamma} & 0 & 0 \\
 \frac{\lambda_G}{2 (\lambda_L + \gamma)} & 0 & 0 &    \frac{\lambda_L}{2 (\lambda_L  + \gamma)} \\
 \frac{\lambda_G}{\lambda_L +  \gamma} & 0 & 0  & \frac{\lambda_L}{\lambda_L  + \gamma} \\
\frac{\lambda_G}{\gamma} & 0 & 0 & 0  \end{pmatrix}. \end{eqnarray}
It is then straightforward to show that
\begin{eqnarray} \label{eq:house:E:3}
\mathbf{M} = (\mathbf{I} - \mathbf{P})^{-1} \Phi = \begin{pmatrix} \frac{\lambda_G}{\gamma} &  \frac{2 \lambda_L}{2 \lambda_L  +\gamma} & 0 & \frac{\lambda_L^2 (\lambda_L + 2 \gamma)}{(2 \lambda_L + \gamma)(\lambda_L + \gamma)^2} \\
 \frac{\lambda_G}{ \gamma} & 0 & 0 &    \frac{\lambda_L (\lambda_L + 2 \gamma)}{2 (\lambda_L  + \gamma)^2} \\
 \frac{\lambda_G}{\gamma} & 0 & 0  & \frac{\lambda_L}{\lambda_L  + \gamma} \\
\frac{\lambda_G}{\gamma} & 0 & 0 & 0  \end{pmatrix} \end{eqnarray}
There are a couple of observations to make concerning $\mathbf{M}$. Firstly, regardless of at what stage of the household epidemic an individual is infected, the mean number of global contacts, and hence, the number of infectious units of type $(h-1,1)$ that are created by the individual is $\lambda_G/\gamma$. Secondly, no individuals of type $(1,1)$ are created in the epidemic since a household can only reach this state from $(1,2)$ and through the recovery of the other infective. More generally, an individual does not start as an infectious unit of type $(a,1)$, where $1 \leq a \leq h-2$, although it is helpful to define such infectious units for the progression of the epidemic.

It follows from \eqref{eq:house:E:3} by removing the redundant row and column for state $(1,1)$ individuals, that the basic reproduction number, $R_0$, solves the cubic equation
\begin{eqnarray} \label{eq:house:E:4}
s^3 - \frac{\lambda_G}{\gamma} s^2 - \frac{\lambda_G}{\gamma} \left\{\frac{2 \lambda_L}{2 \lambda_L  +\gamma} +\frac{\lambda_L^2 (\lambda_L + 2 \gamma)}{(2 \lambda_L + \gamma)(\lambda_L + \gamma)^2} \right\} s -
\frac{\lambda_G}{\gamma} \left\{ \frac{2 \lambda_L}{2 \lambda_L  +\gamma} \frac{\lambda_L (\lambda_L + 2 \gamma)}{2 (\lambda_L  + \gamma)^2}  \right\} =0. \end{eqnarray}
We note that in the notation of \cite{Pellis_etal}, $\mu_G = \lambda_G/\gamma$, $\mu_0 =1$,
\[ \mu_1 = \frac{2 \lambda_L}{2 \lambda_L  +\gamma} +\frac{\lambda_L^2 (\lambda_L + 2 \gamma)}{(2 \lambda_L + \gamma)(\lambda_L + \gamma)^2} \]
and
\[ \mu_2 = \frac{2 \lambda_L}{2 \lambda_L  +\gamma} \frac{\lambda_L (\lambda_L + 2 \gamma)}{2 (\lambda_L  + \gamma)^2}, \] where $\mu_i$ $(i=0,1,\ldots)$ denotes the number of infectives in generation $i$ of the household epidemic, see also \cite{Ball_etal}, Section 3.1.3.
Therefore we can rewrite  \eqref{eq:house:E:4} as
\begin{eqnarray} \label{eq:house:E:5} s^3 - \sum_{i=0}^2 \mu_G \mu_i s^{2-i} = 0, \end{eqnarray}
which is equivalent to \cite{Pellis_etal}, (3.3), and hence obtain an identical $R_0$ to $R_0^g$ defined in \cite{Ball_etal}.

We proceed by showing that for the Markov household epidemic model $R_0$ obtained as the maximal eigenvalue of $\mathbf{M}$ corresponds $R_0^g$ defined in  \cite{Ball_etal} for any $h \geq 1$. In order to do this it is helpful to write
\begin{eqnarray} \label{eq:house:E:6} \mathbf{M} =  \mathbf{G} + \mathbf{U}, \end{eqnarray}
where $\mathbf{G}$ is the $K \times K$ matrix with $G_{k1} = \mu_G$ $(1 \leq k \leq K)$ and $G_{kj} =0$ otherwise. Then $\mathbf{G}$ and $\mathbf{U}$ denote the matrices of global and local  infections, respectively.
For $i=0,1,2,\ldots,h-1$, let $\nu_i = \sum_{j=1}^K u_{1j}^i$, the sum of the first row of $\mathbf{U}^i$. The key observation is that $\nu_i$ denotes the mean number of individuals in generation $i$ of the household epidemic with $\mathbf{U}^0 = \mathbf{I}$, the identity matrix (the initial infective in the household is classed as generation 0) and $\mathbf{U}^i = \mathbf{0}$ for $i \geq h$.

For $0 \leq a,b \leq h-1$, let $y_{(a,b)}^{(n)}$ denote the mean number of type $(a,b)$ individuals in the $n^{th}$ generation of the epidemic process. Then $y_{(h-1,1)}^{(0)}=1$ (the initial infective) and for all $(a,b) \neq (h-1,1)$, $y_{(a,b)}^{(0)}=0$.  Let $\mathbf{y}^{(n)} = (y_{(a,b)}^{(n)})$ denote the mean number of individuals of each type in the $n^{th}$ generation of the epidemic process with the convention that $y_{(h-1,1)}^{(n)}$ is the first entry of $\mathbf{y}^{(n)}$. Then for $n \geq 1$, $\mathbf{y}^{(n)}$ solves
\begin{eqnarray} \label{eq:house:E:7} \mathbf{y}^{(n)} = \mathbf{y}^{(n-1)} \mathbf{M}. \end{eqnarray}
The proof of \eqref{eq:house:E:7} mimics the proof of  \cite{Pellis_etal}, Lemma 2, and it follows by induction that
\begin{eqnarray} \label{eq:house:E:8} \mathbf{y}^{(n)} = \mathbf{y}^{(0)} \mathbf{M}^n. \end{eqnarray}

Let $x_{n,i}$ $(n=0,1,\ldots;i=0,1,\ldots,h-1)$ be defined as in \cite{Pellis_etal}, Lemma 1, with $x_{n,i}$ denoting the mean number of individuals in the $n^{th}$ generation of the epidemic who belong to the $i^{th}$ generation of the household epidemic. We again employ the convention that the initial infective individual in the household represents generation 0. It is shown in \cite{Pellis_etal}, Lemma 1, (3.5) and (3.6) that
\begin{eqnarray} \label{eq:Pellis_etal:1} x_{n,0} = \mu_G \sum_{i=0}^{h-1} x_{n-1,i}, \end{eqnarray}
and
\begin{eqnarray} \label{eq:Pellis_etal:2} x_{n,i} = \mu_i x_{n-i,0}, \end{eqnarray}
where $\mu_i$ is the mean number of infectives in generation $i$ of a household epidemic, $x_{0,0}=1$ and $x_{0,i} =0$ $(i=1,2,\ldots,h-1)$. Let $\mathbf{x}^{(n)} = (x_{n,0}, x_{n,1}, \ldots, x_{n,h-1})$.
${\newtheorem{lemma}{Lemma}[section]}$
\begin{lemma} \label{lem1} For $n=0,1,\ldots$,
\begin{eqnarray} \label{eq:house:E:9} y^{(n)}_{(h-1,1)} = x_{n,0}. \end{eqnarray}

Let $x_n = \mathbf{x}^{(n)} \mathbf{1} = \sum_{j=0}^{h-1} x_{n,j}$ and $y_n = \mathbf{y}^{(n)} \mathbf{1} = \sum_{(a,b)} y^{(n)}_{(a,b)}$, then for $n=0,1,\ldots$,
\begin{eqnarray} \label{eq:house:E:9a} y_n = x_n. \end{eqnarray}
\end{lemma}

Before proving Lemma \ref{lem1}, we prove Lemma \ref{lem2} which gives $\mu_i$ in terms of the local reproduction matrix $\mathbf{U}$.
${\newtheorem{lemma2}[lemma]{Lemma}}$
\begin{lemma2} \label{lem2} For $i=0,1,\ldots,h-1$,
\begin{eqnarray} \label{eq:house:E:10} \mu_i = \sum_{(c,d)} u_{(h-1,1),(c,d)}^i = \nu_i. \end{eqnarray}
\end{lemma2}
{\bf Proof.} Let $Z_{(a,b)}^{(i)}$ denote the total number of individuals of type $(a,b)$ in generation $i$ of a household epidemic. Note that $Z_{(a,b)}^{(i)}$ will be either 0 or 1 and $Z_{(a,b)}^{(i)} =1$ if an infection takes place in a household in state $(a+1,b-1)$ with the infector belonging to generation $i-1$. Then by definition
\begin{eqnarray} \label{eq:house:E:11} \mu_i = \sum_{(a,b)} \ez [ Z_{(a,b)}^{(i)}]. \end{eqnarray}
We note that $Z_{(h-1,1)}^{(0)} =1$ and for $(a,b) \neq (h-1,1)$, $Z_{(a,b)}^{(0)} =0$, giving $\mu_0 =1$. Since $\mathbf{U}^0$ is the identity matrix, we have that
$\sum_{(c,d)} u_{(h-1,1),(c,d)}^0 = 1$ also.

For $i=1,2,\ldots,h-1$, we have that
\begin{eqnarray} \label{eq:house:E:12}  \ez [ Z_{(a,b)}^{(i)}] = \ez[\ez [Z_{(a,b)}^{(i)} | \mathbf{Z}^{(i-1)}] ], \end{eqnarray}
where $\mathbf{Z}^{(i-1)} = (Z_{(a,b)}^{(i-1)})$. Now
\begin{eqnarray} \label{eq:house:E:13}  \ez [Z_{(a,b)}^{(i)} | \mathbf{Z}^{(i-1)}] = \sum_{(c,d)} u_{(c,d),(a,b)} Z_{(c,d)}^{(i-1)}, \end{eqnarray}
since $u_{(c,d),(a,b)}$ is the probability that a type $(c,d)$ individual will infect an individual to create a type $(a,b)$ infective. Therefore
taking expectations of both sides of \eqref{eq:house:E:13} yields
\begin{eqnarray} \label{eq:house:E:14}  \ez [Z_{(a,b)}^{(i)}] = \sum_{(c,d)} u_{(c,d),(a,b)} E[Z_{(c,d)}^{(i-1)}]. \end{eqnarray}

Therefore letting $z_{(a,b)}^{(i)} = \ez [Z_{(a,b)}^{(i)}]$ and $\mathbf{z}^{(i)} = (z_{(a,b)}^{(i)})$ it follows from \eqref{eq:house:E:14} that
\begin{eqnarray} \label{eq:house:E:15}  \mathbf{z}^{(i)} = \mathbf{z}^{(i-1)} \mathbf{U} = \mathbf{z}^{(0)} \mathbf{U}^i. \end{eqnarray}
Hence,
\begin{eqnarray} \label{eq:house:E:16} \mu_i &=& \sum_{(a,b)} z_{(a,b)}^{(i)} \nonumber \\
&=& \sum_{(a,b)} z_{(a,b)}^{(0)}  \sum_{(c,d)} u_{(a,b),(c,d)}^i \nonumber \\
&=&  \sum_{(c,d)} u_{(h-1,1),(c,d)}^i = \nu_i, \end{eqnarray}
as required.
\hfill $\square$

{\bf Proof of Lemma \ref{lem1}.}
We prove the lemma by induction and noting that for $n=0$, $y^{(0)}_{(h-1,1)} = x_{0,0}=1$.

Before proving the inductive step, we note that it follows from \eqref{eq:house:E:7} that
\begin{eqnarray} \label{eq:house:E:17} y_{(h-1,1)}^{(n)} = \frac{\lambda_G}{\gamma} \sum_{(c,d)} y_{(c,d)}^{(n-1)} = \mu_G \sum_{(c,d)} y_{(c,d)}^{(n-1)}\end{eqnarray}
and for $(a,b) \neq (h-1,1)$,
\begin{eqnarray} \label{eq:house:E:18} y_{(a,b)}^{(n)} &=& \sum_{(c,d)} y_{(c,d)}^{(n-1)} u_{(c,d),(a,b)} \nonumber  \\
&=& y_{(h-1,1)}^{(n-1)} u_{(h-1,1),(a,b)} + \sum_{(c,d) \neq (h-1,1)} y_{(c,d)}^{(n-1)} u_{(c,d),(a,b)} \nonumber \\
&=& y_{(h-1,1)}^{(n-1)} u_{(h-1,1),(a,b)} + \sum_{(c,d) \neq (h-1,1)} \left\{ \sum_{(e,f)} y_{(e,f)}^{(n-2)} u_{(e,f),(c,d)} \right\} u_{(c,d),(a,b)} \nonumber \\
&=& y_{(h-1,1)}^{(n-1)} u_{(h-1,1),(a,b)} + \sum_{(e,f)} y_{(e,f)}^{(n-2)} \sum_{(c,d) \neq (h-1,1)} u_{(e,f),(c,d)} u_{(c,d),(a,b)} \nonumber \\
&=& y_{(h-1,1)}^{(n-1)} u_{(h-1,1),(a,b)} + \sum_{(e,f)} y_{(e,f)}^{(n-2)} u_{(e,f),(a,b)}^2. \end{eqnarray}
The final line of \eqref{eq:house:E:18} follows from $u_{(e,f),(h-1)} =0$ for all $(e,f)$.
Then by a simple recursion it follows from \eqref{eq:house:E:18}, after at most $h-1$ steps, that, for $(a,b) \neq (h-1,1)$,
\begin{eqnarray} \label{eq:house:E:19} y_{(a,b)}^{(n)} &=& \sum_{j=1}^{h-1} y_{(h-1,1)}^{(n-j)} u_{(h-1,1),(a,b)}^j. \end{eqnarray}
Note that \eqref{eq:house:E:19} can easily be extended to include $(a,b) = (h-1,1)$ giving
\begin{eqnarray} \label{eq:house:E:20} y_{(a,b)}^{(n)} &=& \sum_{j=0}^{h-1} y_{(h-1,1)}^{(n-j)} u_{(h-1,1),(a,b)}^j. \end{eqnarray}

For $n \geq 1$, we assume the inductive hypothesis that for $0 \leq k \leq n-1$, $y^{(k)}_{(h-1,1)} = x_{k,0}$. Then from \eqref{eq:house:E:20}, we have that
\begin{eqnarray} \label{eq:house:E:21}
y^{(n)}_{(h-1,1)} &=& \sum_{(a,b)} m_{(a,b),(h-1,1)} y^{(n-1)}_{(a,b)}  \nonumber \\
&=& \mu_G \sum_{(a,b)} y^{(n-1)}_{(a,b)} \nonumber \\
&=& \mu_G \sum_{(a,b)} \left\{ \sum_{j=0}^{h-1} y_{(h-1,1)}^{(n-1-j)} u_{(h-1,1),(a,b)}^j \right\}  \nonumber \\
&=& \mu_G \sum_{j=0}^{h-1} y_{(h-1,1)}^{(n-1-j)} \left( \sum_{(a,b)}u_{(h-1,1),(a,b)}^j  \right). \end{eqnarray}
Using the inductive hypothesis and Lemma \ref{lem2}, we have from \eqref{eq:house:E:21} that
\begin{eqnarray} \label{eq:house:E:22}
y^{(n)}_{(h-1,1)}&=& \mu_G \sum_{j=0}^{h-1} x_{(n-1-j),0} \mu_j =x_{n,0},  \end{eqnarray}
as required for \eqref{eq:house:E:9}.

Using a similar line of argument,
\begin{eqnarray} \label{eq:house:E:22a}
y_n = \mathbf{y}^{(n)} \mathbf{1} &=& \sum_{(a,b)} y^{(n)}_{(a,b)} \nonumber \\
&=& \sum_{(a,b)} \left\{ \sum_{j=0}^{h-1} y^{(n-j)}_{(h-1,1)} u_{(h-1,1),(a,b)}^j \right\} \nonumber \\
&=& \sum_{j=0}^{h-1} y^{(n-j)}_{(h-1,1)} \sum_{(a,b)} \left\{ u_{(h-1,1),(a,b)}^j \right\} \nonumber \\
&=& \sum_{j=0}^{h-1} x_{n-j,0} \mu_j = x_n,  \end{eqnarray}
as required for \eqref{eq:house:E:9a}. \hfill $\square$

Therefore we have shown that the two representations of the household epidemic given in \cite{Pellis_etal} and in this paper give the same mean number of infectives and the same mean number of new household epidemics in generation $n$ $(n=0,1,\ldots)$. This is a key component in showing that $\mathbf{M}$ and $\mathbf{A}$, the mean reproductive matrix given in \cite{Pellis_etal} by
 \begin{eqnarray} \label{eq:house:E:23}
\mathbf{A} = \begin{pmatrix} \mu_G \mu_0 & 1 & 0 & \cdots & 0 \\
\mu_G \mu_1 & 0 & 1 & & 0 \\
\vdots & && \ddots & 0 \\
\mu_G \mu_{h-2} & 0 & 0 & & 1 \\
\mu_G \mu_{h-1} & 0 & 0& \cdots & 0 \\  \end{pmatrix} \end{eqnarray} have the same largest eigenvalue.

Let $\rho_A$ and $\rho_M$ denote the largest eigenvalues of $\mathbf{A}$ and $\mathbf{M}$, respectively.
Let $\mathbf{z}_L$ and $\mathbf{z}_R$ denote the normalised left and right eigenvectors corresponding to $\rho_A$ with $\mathbf{z}_L \mathbf{z}_R = 1$.
In \cite{Pellis_etal}, Lemma 3, it is note that
\begin{eqnarray} \label{eq:house:E:24} \mathbf{A} = \rho_A \mathbf{C}_A + B_A, \end{eqnarray}
where $\mathbf{C}_A = \mathbf{z}_R \mathbf{z}_L$ and $\rho_A^{-n} B_A^n \rightarrow \mathbf{0}$ as $n \rightarrow \infty$. This implies that if $x_n = \mathbf{x}^{(n)} \mathbf{1}$, the mean number of individuals infected in the $n^{th}$ generation of the epidemic then
\begin{eqnarray} \label{eq:house:E:25} (y_n^{1/n} =) x_n^{1/n} \rightarrow \rho_A \hspace{0.5cm} \mbox{as } n \rightarrow \infty. \end{eqnarray}

As observed earlier the construction of $\mathbf{M}$ results in $\mathbf{0}$ columns corresponding to infectious units which can arise through the removal of an infective. Let $\tilde{\mathbf{M}}$ denote the matrix obtained by removing the $\mathbf{0}$ columns and corresponding rows from $\mathbf{M}$. The eigenvalues of $\mathbf{M}$ will consist of the eigenvalues of $\tilde{\mathbf{M}}$ plus repeated 0 eigenvalues, one for each $\mathbf{0}$  column.
Let $\mathbf{w}_L$ and $\mathbf{w}_R$ denote the normalised left and right eigenvectors corresponding to $\rho_{\tilde{M}}$ with $\mathbf{w}_L \mathbf{w}_R = 1$. Then since $\tilde{\mathbf{M}}$ is a positively regular matrix by the Perron-Frobenius theorem, $\tilde{\mathbf{M}}$ (and hence $\mathbf{M}$) has a unique real and positive largest eigenvalue, $\rho_M$. Moreover,
\begin{eqnarray} \label{eq:house:E:26} \tilde{\mathbf{M}} = \rho_M \mathbf{C}_M + B_M, \end{eqnarray}
where $\mathbf{C}_M = \mathbf{w}_R \mathbf{w}_L$ and $\rho_M^{-n} B_M^n \rightarrow \mathbf{0}$ as $n \rightarrow \infty$.
 Then following the arguments in the proof of \cite{Pellis_etal}, Lemma 3,
\begin{eqnarray} \label{eq:house:E:27} y_n^{1/n} \rightarrow \rho_M \hspace{0.5cm} \mbox{as } n \rightarrow \infty. \end{eqnarray}
Since $x_n = y_n$ $(n=0,1,2,\ldots)$, it follows from \eqref{eq:house:E:25} and \eqref{eq:house:E:27} that $\rho_M = \rho_A$ and therefore that the two constructions of the epidemic process give the same basic reproduction number $R_0$. 

%
%

\subsection{$SIS$ sexually transmitted disease model} \label{S:example:sex}

We begin by describing the $SIS$ sexually transmitted disease model which provided the motivation for this work and then study the construction of $\mathbf{M}$.

\subsubsection{Model}

We consider a model for a population of sexually active individuals who alternate between being in a relationship and being single. For simplicity of presentation, we assume a homosexual model where each relationship comprises of two individuals. The extension to a heterosexual population with equal numbers of males and females is straightforward.
We assume $SIS$ disease dynamics with infectious individuals returning to the susceptible state on recovery from the disease. There are two key dynamics underlying the spread of the disease. The formation and dissolution of relationships and the transmission of the disease.

%
%

Individuals are termed as either single (not currently in a relationship) or coupled (currently in a relationship).
We assume that each single individual seeks to instigate the formation of relationship at the points of a homogeneous Poisson point process with rate $\alpha/2$ with the individual with whom they seek to form a relationship 
chosen uniformly at random from the population. (The rate $\alpha/2$ allows for individuals to be both instigators and contacted individuals.) If a contacted individual is single, they agrees to form a relationship with the instigator, otherwise the individual is already in a relationship and remains with their current partner.  The lifetimes of relationships are independent and identically distributed according to a non-negative random variable $T$ with mean $1/\delta$. For a Markovian model we take $T \sim {\rm Exp} (\delta)$ corresponding to relationships dissolving at rate $\delta$. When a relationship dissolves the individuals involved return to the single state.  Therefore there is a constant flux of individuals going from single to coupled and back again. We assume that the disease is introduced into the population at time $t=0$ with the population in stationarity with regards relationship status. The proportion, $\sigma$, of the population who are single in stationarity is given in \cite{Lashari_Trapman} with
\begin{eqnarray} \label{eq:model:2}
\sigma^2 \alpha &=& \delta (1 -\sigma) \nonumber \\
\sigma &=& \frac{- \delta + \sqrt{\delta^2 + 4 \delta \alpha}}{2 \alpha}.
\end{eqnarray} Thus $\tilde{\alpha} = \alpha \sigma$ is the rate at which a single individual enters a relationship.
We assume that the relationship dynamics are in a steady state when the disease is introduced and the introduction of the diseases does not affect relationship dynamics.

%
%
We now turn to the disease dynamics. We assume $SIS$ dynamics, in that individuals alternate between being susceptible and infectious and on recovery from being infectious an individual immediately reenters the susceptible state. We allow for two types of sexual contacts, those within relationships and {\it casual} contacts which occur outside relationships. The casual contacts which we term {\it one-night stands} represent single sexual encounters capturing short term liaisons. We assume that the infectious periods  are independent and identically distributed according  to a non-negative random variable $Q$, where $Q \sim {\rm Exp} (\gamma)$ for a Markovian model.
 Whilst in a relationship, we assume that an infectious individual makes infectious contact with their partner at the points of a homogeneous Poisson point process with rate $\beta$. We assume that individuals can also partake in, and transmit the disease via, one-off sexual contacts (one-night stands). We assume that individuals in relationships are less likely to take part in a one-night stand with probability $\rho$ of having a one-night stand. Therefore
we assume that a single individual (individual in a relationship) seeks to make infectious contact via one-night stands at the points of a homogeneous Poisson point process with rate $\omega$ $(\rho \omega)$, where $\omega$ amalgamates the propensity for partaking in a one-night stand with the transmissibility of the disease during a one-night stand. If an individual attempts to have a one-night stand with somebody in a relationship, there is only probability $\rho$ of the one-night stand occurring. Thus $\rho=0$ denotes that individuals in relationships are faithful, whilst $\rho =1$ denotes that there is no difference between those in or out of a relationship with regards one-night stands.

In the early stages of the epidemic with high probability a single infective will form a relationship with a susceptible individual and one-night stands with an individual in a relationship will be with a totally susceptible relationship.

\subsubsection{Construction of $\mathbf{M}$}

For this model there are three types of infectious units; single infective, couple with one infective and couple with two infectives which we will term types 1, 2 and 3, respectively. The possible events and their rates of occurring are presented in Table \ref{tab:sex:1}.

\begin{table}
\begin{tabular}{l|ccc}
Event Type & Single infective & \multicolumn{2}{c}{Infective in a relationship} \\
& & Susceptible Partner & Infectious partner \\
\hline
Relationship form & $\alpha \sigma$ &  --   & -- \\
Relationship dissolve  &  -- & $\delta$ & $\delta$ \\
One-night stand single & $\omega \sigma$ & $\rho \omega \sigma$ & $\rho \omega \sigma$ \\
One-night stand relationship & $\rho \omega \sigma$ & $\rho^2 \omega \sigma$ & $\rho^2 \omega \sigma$ \\
Infect partner & -- & $\beta$ & -- \\
Partner recovers & -- & -- & $\gamma$ \\
Recovers & $\gamma$ & $\gamma$ & $\gamma$ \\
\end{tabular}
\caption{Events and their rates of occurring for an infectious individual in each type of infectious unit.} \label{tab:sex:1}
\end{table}


It is straightforward from Table \ref{tab:sex:1} to construct $\Phi^E$ and $\mathbf{P}^E$ in terms of the next event to occur. For $\Phi^E$, the next event will create at most one infective and we only need to compute the probability of each type of infection. Hence,
\begin{eqnarray} \label{eq:sex:1}
\Phi^E = \begin{pmatrix} \frac{\omega \sigma}{\alpha \sigma + \omega \{1- (1-\rho) (1-\sigma)\} + \gamma} & \frac{\rho \omega \sigma}{\alpha \sigma + \omega \{1- (1-\rho) (1-\sigma)\} + \gamma} & 0 \\
\frac{\rho \omega \sigma}{\delta + \rho\omega \{1- (1-\rho) (1-\sigma)\} + \beta + \gamma} & \frac{\rho^2 \omega \sigma}{\delta + \rho\omega \{1- (1-\rho) (1-\sigma)\} + \beta + \gamma} & \frac{\beta}{\delta + \rho\omega \{1- (1-\rho) (1-\sigma)\} + \beta + \gamma} \\
\frac{\rho\omega \sigma}{\delta + \rho \omega \{1- (1-\rho) (1-\sigma)\} + 2 \gamma} & \frac{\rho^2 \omega \sigma}{\delta + \rho \omega \{1- (1-\rho) (1-\sigma)\} + 2 \gamma} & 0  \end{pmatrix}. \end{eqnarray}
Similarly by considering the transition at each event, we have that
\begin{eqnarray} \label{eq:sex:2}
\mathbf{P}^E = \begin{pmatrix} \frac{\omega \{1- (1-\rho) (1-\sigma)\}}{\alpha \sigma + \omega \{1- (1-\rho) (1-\sigma)\} + \gamma} & \frac{\alpha \sigma}{\alpha \sigma + \omega \{1- (1-\rho) (1-\sigma)\} + \gamma} & 0 \\
\frac{\delta}{\delta + \rho\omega \{1- (1-\rho) (1-\sigma)\} + \beta + \gamma} & \frac{\rho\omega \{1- (1-\rho) (1-\sigma)\}}{\delta + \rho\omega \{1- (1-\rho) (1-\sigma)\} + \beta + \gamma} & \frac{\beta}{\delta + \rho\omega \{1- (1-\rho) (1-\sigma)\} + \beta + \gamma} \\
\frac{\delta}{\delta + \rho \omega \{1- (1-\rho) (1-\sigma)\} + 2 \gamma} & \frac{\gamma}{\delta + \rho \omega \{1- (1-\rho) (1-\sigma)\} + 2 \gamma} & \frac{\rho\omega \{1- (1-\rho) (1-\sigma)\}}{\delta + \rho \omega \{1- (1-\rho) (1-\sigma)\} + 2 \gamma}  \end{pmatrix}. \end{eqnarray}

One-night stands do not alter the relationship and hence do not constitute transition events. Given that the number of one-night stands made by an infectious individual in an interval of a given length follows a Poisson distribution with mean proportional to the length of the interval it is straightforward to show that
\begin{eqnarray} \label{eq:sex:3}
\Phi = \begin{pmatrix} \frac{\omega \sigma}{\alpha \sigma +  \gamma} & \frac{\rho \omega \sigma}{\alpha \sigma + \gamma} & 0 \\
\frac{\rho \omega \sigma}{\delta  + \beta + \gamma} & \frac{\rho^2 \omega \sigma}{\delta + \beta + \gamma} & \frac{\beta}{\delta + \beta + \gamma} \\
\frac{\rho\omega \sigma}{\delta  + 2 \gamma} & \frac{\rho^2 \omega \sigma}{\delta + 2 \gamma} & 0  \end{pmatrix}, \end{eqnarray}
and that the transition matrix is given by
\begin{eqnarray} \label{eq:sex:4}
\mathbf{P} = \begin{pmatrix} 0 & \frac{\alpha \sigma}{\alpha \sigma +  \gamma} & 0 \\
\frac{\delta}{\delta + \beta + \gamma} & 0 & \frac{\beta}{\delta + \beta + \gamma} \\
\frac{\delta}{\delta + 2 \gamma} & \frac{\gamma}{\delta + 2 \gamma} & 0  \end{pmatrix}. \end{eqnarray}
Straightforward, but tedious algebra, gives that
\begin{eqnarray} \label{eq:sex:5}
\mathbf{M} &=& (\mathbf{I} - \mathbf{P})^{-1}\Phi =(\mathbf{I} - \mathbf{P}^E)^{-1}\Phi^E \nonumber \\ &=&  \begin{pmatrix} \frac{\sigma \omega (\alpha \sigma \rho + \delta + \gamma)}{(\alpha \sigma + \delta + \gamma) \gamma} &  \frac{\sigma \omega \rho (\alpha \sigma \rho + \delta + \gamma)}{(\alpha \sigma + \delta + \gamma) \gamma} & \frac{\alpha \sigma \beta (\delta + 2 \gamma)}{\gamma (\alpha \sigma + \delta + \gamma)(\beta+ \delta +2 \gamma)} \\
\frac{\sigma \omega (\alpha \sigma \rho + \delta + \gamma \rho)}{(\alpha \sigma + \delta + \gamma) \gamma} &  \frac{\sigma \omega \rho (\alpha \sigma \rho + \delta + \gamma \rho)}{(\alpha \sigma + \delta + \gamma) \gamma} & \frac{ \beta (\delta + 2 \gamma) (\alpha \sigma + \gamma)}{\gamma (\alpha \sigma + \delta + \gamma)(\beta+ \delta +2 \gamma)} \\
\frac{\sigma \omega (\alpha \sigma \rho + \delta + \gamma \rho)}{(\alpha \sigma + \delta + \gamma) \gamma} &  \frac{\sigma \omega \rho (\alpha \sigma \rho + \delta + \gamma \rho)}{(\alpha \sigma + \delta + \gamma) \gamma} & \frac{ \beta \{ \alpha \sigma (\delta + \gamma) +\gamma^2\}}{\gamma (\alpha \sigma + \delta + \gamma)(\beta+ \delta +2 \gamma)} \end{pmatrix}. \end{eqnarray}
Note that the mean number of one-night stands is the same for all individuals who start their infectious period in a relationship, regardless of the infectious status of their partner.
The eigenvalues of $\mathbf{M}$ can be obtained from solving the characteristic equation ${\rm det} (\mathbf{M} - s \mathbf{I}) = 0$, a cubic polynomial in $s$. The resulting algebraic expressions are not very illuminating about $R_0$ and its properties. However, this does allow for simple computation of $R_0$ for specified parameter values.

In the special case $\rho =0$ where only single individuals can have one-night stands, we note that individuals can only enter the infectious state as a member of an infectious unit of type 1 or 3. Furthermore, if $\omega =0$, there are no one-night stands and individuals only become infected via an infectious partner within a relationship. In this case the first two columns of $\mathbf{M}$ become $\mathbf{0}$ and
\begin{eqnarray} \label{eq:sex:6}
R_0 = M_{3,3} =\frac{ \beta \{ \alpha \sigma (\delta + \gamma) +\gamma^2\}}{\gamma (\alpha \sigma + \delta + \gamma)(\beta+ \delta +2 \gamma)}, \end{eqnarray}
the mean number of times an individual will successfully infect a partner whilst infectious.

The expression for $R_0$ given in \eqref{eq:sex:6} is very similar to that given in \cite{LKD15}, (30). The only difference for $\omega=0$ between our model and the $SI$ sexually transmitted disease model presented in \cite{LKD15} and \cite{Lashari_Trapman} for individuals with a maximum of one sexual partner is that the model of \cite{LKD15} replaces recovery ($\gamma$) by death ($\mu$) which results in the relationship ending as well as removal of the infective. The model of \cite{LKD15} and \cite{Lashari_Trapman} incorporates birth of susceptibles at rate $N \mu$ to maintain the population size of $O(N)$.

\subsection{Great Circle Epidemic model} \label{S:example:gcm}

The final example we consider in this paper is the great circle $SIR$ epidemic model, see \cite{Ball_Neal03} and references therein. The model assumes that the population consists of $N$ individuals who are equally spaced on the circumference of a circle with the individuals labeled sequentially from 1 to $N$ and individuals 1 and $N$ are neighbours. Thus individuals $i \pm 1 (mod \, N)$ are the neighbours of individual $i$ $(i=1,2,\ldots,N)$. Individuals, whilst infectious, make both local and global infectious contacts as in the household model. An individual makes global infectious contacts at the points of a homogeneous Poisson point process with rate $\lambda_G$ with the individual contacted chosen uniformly at random from the entire population. An individual makes local infectious contacts with a given neighbour at the points of a homogeneous Poisson point process with rate $ \lambda_L$. Finally, the infectious periods are independently and exponentially distributed with mean $1/\gamma$.

An infectious individual in the great circle model can be characterised by the total number of susceptible neighbours that it has which can be $2, 1$ or 0. In the initial stages of the epidemic with $N$ large, with high probability, an individual infected globally will initially have 2 susceptible neighbours, whereas an individual infected locally will, with high probability, have 1 susceptible neighbour when they are infected. An infective with $k$ $(k=0, 1,2)$ susceptible neighbours makes ${\rm Po} (\lambda_G/\{k \lambda_L + \gamma \})$ global infectious contacts before a local infection or recovery event with the probability that the event is the infection of a neighbour $k \lambda_L /(k \lambda_L + \gamma)$. Therefore if we construct $\Phi$ and $\mathbf{P}$ in terms of descending number of susceptible neighbours we have that
\begin{eqnarray} \label{eq:gcm:1}
\Phi = \begin{pmatrix} \frac{\lambda_G}{2 \lambda_L + \gamma} & \frac{2 \lambda_L}{2 \lambda_L + \gamma} & 0 \\
\frac{\lambda_G}{ \lambda_L + \gamma} & 0 & \frac{ \lambda_L}{ \lambda_L + \gamma}  \\ \frac{\lambda_G}{\gamma}  & 0& 0  \end{pmatrix}, \end{eqnarray}
and
\begin{eqnarray} \label{eq:gcm:2}
\mathbf{P} = \begin{pmatrix} 0 & \frac{2 \lambda_L}{2 \lambda_L + \gamma} & 0 \\
0 & 0 & \frac{\lambda_L}{\lambda_L + \gamma} \\
0 & 0 & 0  \end{pmatrix}. \end{eqnarray}
It is then straightforward to show that
\begin{eqnarray} \label{eq:gcm:3}
\mathbf{M} = \begin{pmatrix} \frac{\lambda_G}{\gamma} & \frac{2 \lambda_L}{\lambda_L +\gamma} & 0 \\
\frac{\lambda_G}{\gamma}   & \frac{\lambda_L}{\lambda_L +\gamma} & 0 \\
\frac{\lambda_G}{\gamma}  & 0 & 0  \end{pmatrix}. \end{eqnarray}
We observe that no individuals are created with 0 susceptible neighbours and we only need to consider the mean offspring distributions for type 1 and type 2 infectives. This gives $R_0$ as the solution of the quadratic equation,
\begin{eqnarray} \label{eq:gcm:4} \left( \frac{\lambda_G}{\gamma} - s \right) \left( \frac{\lambda_L}{\lambda_L +\gamma} - s \right) - \frac{\lambda_G}{\gamma} \times \frac{2 \lambda_L}{\lambda_L +\gamma} &=&0 \nonumber \\
s^2 - (\mu_G + p_L) s - \mu_G p_L &=& 0,
\end{eqnarray} where $\mu_G = \lambda_G/\gamma$ denotes the mean number of global infectious contacts made by an infective and $p_L = \lambda_L /(\lambda_L + \gamma)$ denotes the probability an infective makes infects a given susceptible neighbour. This yields
 \begin{eqnarray} \label{eq:gcm:5} R_0 = \frac{p_L + \mu_G + \sqrt{(p_L + \mu_G)^2 + 4 p_L \mu_G}}{2}. \end{eqnarray}
In \cite{BMST}, \cite{Ball_Neal02} and \cite{Ball_Neal03}, the threshold parameters $R_\ast$ is defined for the great circle model as the mean number of global infectious contacts emanating from a local infectious clump, where a local infectious clump is defined to be the epidemic generated by a single infective by only considering local (neighbour) infectious contacts. From \cite{Ball_Neal02}, (3.12),
 \begin{eqnarray} \label{eq:gcm:6} R_\ast = \mu_G \frac{1 + p_L}{1- p_L}. \end{eqnarray} It is trivial to show that $R_0 =1$ $(R_0 <1; R_0 >1)$ if and only if $R_\ast =1$ $(R_\ast <1; R_\ast >1)$ confirming $R_0$ and $R_\ast$ as equivalent threshold parameters for the epidemic model.

In contrast to the household $SIR$ epidemic model (Section \ref{S:example:house}) and the $SIS$ sexually transmitted disease (Section \ref{S:example:sex}) for the great circle model it is trivial to extend the above definition of $R_0$ to a general infectious period distribution $T$. Let $\mu_T = \ez [T]$, the mean of the infectious period and $\phi_T (\theta) = \ez [ \exp(- \theta T)]$ $(\theta \in \mathbb{R}^+)$, the Laplace transform of the infectious period. Thus $\mu_G$ and $p_L$ become $\lambda_G \mu_T$ and $1 - \phi (\lambda_L)$, respectively. Then the probability that a globally infected individual infects 0, 1 or 2 of its initially susceptible neighbours is $\phi (2 \lambda_L)$, $2  \{ \phi ( \lambda_L) - \phi (2 \lambda_L) \}$ and $1 - 2 \phi ( \lambda_L) + \phi (2 \lambda_L) $, respectively. Similarly the probability that a locally infected individual infects its initially susceptible neighbour is $p_L =1- \phi (\lambda_L)$. Since the mean number of global infectious contacts made by an infective is $\mu_G (= \lambda_G \mu_T)$ regardless of whether or not the individual is infected globally or locally, we can derive directly the mean offspring matrix $\mathbf{M}$ is terms of those infected globally (initially 2 susceptible neighbours) and those infected locally (initially 1 susceptible neighbour) with
 \begin{eqnarray} \label{eq:gcm:7}
\mathbf{M} = \begin{pmatrix} \mu_G & 2 p_L \\
\mu_G & p_L \end{pmatrix}. \end{eqnarray} Therefore after omitting the final column (and row) of $\mathbf{M}$ from \eqref{eq:gcm:3}, the equation for $\mathbf{M}$ is identical in \eqref{eq:gcm:3} and  \eqref{eq:gcm:7}, and hence \eqref{eq:gcm:5} holds for $R_0$ for a general infectious period distribution $T$.  

\section{Conclusions} \label{S:conc}

In this paper we have given a simple definition and derivation of $R_0$ for structured populations by considering the population to consist of different types of infectious units. This multi-type approach to constructing $R_0$, via the mean offspring matrix of the different types of infectious units, follows a widely established mechanism introduced in \cite{Diekmann_etal}. Moreover, we have demonstrated that for $SIR$ household epidemic models that $R_0$ coincides with $R_0^g$, the generational basic reproduction number defined in \cite{Ball_etal}. In \cite{Ball_etal}, the rank generational basic reproduction number, $R_0^r$, is also considered and is taken to be the default choice of $R_0$ in that paper.

For the household $SIR$ epidemic model is straightforward to define and construct a rank generational mean reproduction matrix $\mathbf{M}^R$ for a general infectious period distribution, $T$. The approach is to represent the evolution of the epidemic as a discrete time process generation-by-generation. This approach ignores the time dynamics of the epidemic but leaves the final size unaffected and dates back to \cite{Ludwig75}.
 The initial infective in the household (infected by a global infection) forms generation 0. Then for $i=1,2,\ldots,h-1$, the infectious contacts by the infectives in generation $i-1$ are considered and any susceptible individual contacted by an infective in generation $i-1$ will become an infective in generation $i$. We define an infective to be a type $(a,b)$ individual if the generation of the household epidemic in which they are an infective has $b$ infectives and $a$ susceptibles. In the construction of $\mathbf{M}^R$, we again look at the mean number of infections attributable to a given infective with details provided in Appendix \ref{App:rank}.  Let $\mu_i^R$ denote the mean number of infectives in generation $i$ of the household epidemic then it is shown in \cite{Ball_etal}, Section 3.1.3, that for all $k \geq 1$, $\sum_{i=0}^k \mu_i^R \geq  \sum_{i=0}^k \mu_i$, which in turn implies $R_0^r \geq R_0^g$. The construction of $\mathbf{M}^R$ is straightforward using \cite{Pellis_etal}, Appendix A and we provide a brief outline in Appendix \ref{App:rank} how similar arguments to those used in Section \ref{S:example:house} can be used to show that $R_0^r$ is the maximal eigenvalue of $\mathbf{M}^R$.

The rank generational construction is natural for the $SIR$ epidemic model and allows us to move beyond $T \sim {\rm Exp} (\gamma)$ but does not readily apply to $SIS$ epidemic models. Extensions of the $SIS$ epidemic model are possible by using the method of stages, see \cite{Barbour76}, where $T$ can be expressed as a sum or mixture of exponential distributions and by extending the number of infectious units to allow for individuals in different stages of the infectious period. In principle $\mathbf{P}$ and $\Phi$ can be constructed as above but the number of possible infectious units rapidly grows making the calculations more cumbersome.

\section*{Acknowledgements}

TT was supported by a PhD scholarship, grant number ST\_2965 Lancaster U, from the Royal Thai Office
of Education Affairs.

\appendix

\section{Construction of $\mathbf{M}^R$} \label{App:rank}

We give a brief outline of the construction of $\mathbf{M}^R$ and demonstrate that its maximal eigenvalue coincides with $R_0^r$ given in \cite{Ball_etal}, Section 3.1.3.

We begin by computing the transition probabilities for a household epidemic in state $(a,b)$, $a$ susceptibles and $b$ infectives. By considering the total amount of infection, $I_b$, generated by the $b$ infectives and the fact that the infectious periods, $T$, are independent and identically distributed, it was shown in \cite{Pellis_etal}, Appendix A, that $X_{(a,b)} | I_b \sim {\rm Bin} (a, 1- \exp(-\lambda_L I_b))$ with
\begin{eqnarray} \label{eq:rank:3} \pz (X_{(a,b)} = c ) &=& \ez[ \pz (X_{(a,b)} = c | I_b)] \nonumber \\
&=& \binom{a}{c} \ez \left[ \{1 -\exp(-\lambda_L I_b) \}^c \exp(-\lambda_L I_b)^{a-c} \right] \nonumber \\
&=& \binom{a}{c} \sum_{j=0}^c \binom{c}{j} (-1)^j \ez [ \exp(- \lambda_L (a+j-c) I_b] \nonumber \\
&=& \binom{a}{c} \sum_{j=0}^c \binom{c}{j} (-1)^j \phi_T (\lambda_L (a+j-c))^b, \end{eqnarray} where $\phi_T (\theta) = \ez [ \exp(-\theta T)]$ is the Laplace transform of the infectious period distribution.
Given that if a household epidemic transitions from state $(a,b)$ to state $(a-c,c)$, the mean number of infections due to any given infective is simply $c/b$.


For the rank generational representation of the epidemic, we can again subsume all states $(0,b)$ into $(0,1)^\ast$. We note that in contrast to Section \ref{S:example:house}, epidemic states $(a,1)$ $(1 \leq a \leq h-2)$ can arise from the epidemic process whilst states $(h-b,b)$ $(b>1)$ will not occur.
For all $\{(a,b); b >0, a+b \leq h\}$, we have that $M_{(a,b),(h-1,1)}^R = \mu_G= \lambda_G \ez[T]$ (the mean number of global infectious contacts made by an individual) and for $(d,c) \neq (h-1,1)$,
\begin{eqnarray} \label{eq:rank:4} M_{(a,b),(d,c)}^R = \left\{ \begin{array}{ll} \frac{c}{b} \binom{a}{c} \sum_{j=0}^c \binom{c}{j} (-1)^j \phi_T (\lambda_L (a+j-c))^b & \mbox{if } d = a-c >0 \\
\frac{a}{b} \sum_{j=0}^a \binom{a}{j} (-1)^j \phi_T (\lambda_L j)^b & \mbox{if } (d,c)=(0,1)^\ast  \\
0 &  \mbox{otherwise}. \end{array} \right. \end{eqnarray}
We can proceed along identical lines to \eqref{eq:house:E:6} in decomposing $\mathbf{M}^R$ into
\begin{eqnarray} \label{eq:rank:4a} \mathbf{M}^R = \mathbf{G} + \mathbf{U}^R, \end{eqnarray} where $\mathbf{G}$ is the $K \times K$ matrix ($K$ denotes the total number of infectious states) with $G_{k1} = \mu_G$ $(1 \leq k \leq K)$ and $G_{kj}=0$ otherwise. For $i=0,1, \ldots, h-1$, let $\mu_i^R$ denote the mean number of individuals in the $i^{th}$ generation of the rank construction of the epidemic, then we have that $\mu_i^R = \sum_{j=1}^K (u_{ij}^R)^i$, the sum of the first row of $(\mathbf{U}^R)^i$. We can follow essentially identical arguments to those given in Section \ref{S:example:house} to show that $R_0^r$ is the maximal eigenvalue of $\mathbf{M}^R$.

To illustrate, this approach we consider households of size $h=3$.  The possible infectious units are $(2,1), (1,1)$ and $(0,1)^\ast$ with mean reproductive matrix
\begin{eqnarray} \label{eq:rank:5} \mathbf{M}^R = \begin{pmatrix} \mu_G  & 2 \{ \phi_T (\lambda_L) - \phi_T (2 \lambda_L) \} & 2 \{ 1- 2 \phi_T (\lambda_L) + \phi_T (2 \lambda_L) \} \\
\mu_G    & 0& 1 - \phi_T (\lambda_L) \\
\mu_G   & 0 & 0  \end{pmatrix}. \end{eqnarray}
The eigenvalues of $\mathbf{M}^R$ are solutions of the cubic equation
\begin{eqnarray} \label{eq:rank:6} s^3 - \mu_G  s^2 - 2 \left\{1 - \phi_T ( \lambda_L) \right\} \mu_G  s - \mu_G \ 2 \{1 - \phi_T (\lambda_L)\}\{ \phi_T (\lambda_L) - \phi_T (2 \lambda_L) \} &=&0 \nonumber \\
s^3 - \mu_G \mu_0^R  s^2 -  \mu_G \mu_1^R  s - \mu_G \mu_2^R &=&0, \end{eqnarray}
where $\mu_0^R =1$, $\mu_1^R= 2 \{ 1 - \phi_T (\lambda_L) \}$ and
$\mu_2^R = 2 \{\phi (\lambda_L) - \phi (2 \lambda_L) \} \{1 - \phi_T (\lambda_L) \}$ are the mean number of infectives in rank generations 0, 1 and 2, respectively of the household epidemic model. Given that \eqref{eq:rank:6} is equivalent to \cite{Pellis_etal}, (3.3), it follows that that maximal eigenvalue of $\mathbf{M}^R$ is $R_0^r$.

\end{document}